\documentstyle[prl,epsf,aps,floats]{revtex}  
\input psfig
\def\met{\mbox{${\hbox{$E$\kern-0.6em\lower-.1ex\hbox{/}}}_T$}} 
\def\mex{\mbox{${\hbox{$E$\kern-0.6em\lower-.1ex\hbox{/}}}_x$}} 
\def\mey{\mbox{${\hbox{$E$\kern-0.6em\lower-.1ex\hbox{/}}}_y$}} 
\def\mexy{\mbox{${\hbox{$E$\kern-0.6em\lower-.1ex\hbox{/}}}_{x,y}$}} 
 
\def\D0{D\O}                            
\begin{document}
\lefthyphenmin=2
\righthyphenmin=3

%
%
\title{
Search for Leptoquark Pairs Decaying to $\boldmath \nu\nu$ + jets in 
$\boldmath p\overline{p}~$ Collisions at $\boldmath \sqrt{s}$=1.8 TeV\\
}

%
\author{                                                                      
V.M.~Abazov,$^{23}$                                                           
B.~Abbott,$^{57}$                                                             
A.~Abdesselam,$^{11}$                                                         
M.~Abolins,$^{50}$                                                            
V.~Abramov,$^{26}$                                                            
B.S.~Acharya,$^{17}$                                                          
D.L.~Adams,$^{59}$                                                            
M.~Adams,$^{37}$                                                              
S.N.~Ahmed,$^{21}$                                                            
G.D.~Alexeev,$^{23}$
% ...
\\                                                                            
\vskip 0.30cm                                                                 
\centerline{(D\O\ Collaboration)}                                             
\vskip 0.30cm                                                                 
}                                                                             
\address{                                                                     
\centerline{$^{1}$Universidad de Buenos Aires, Buenos Aires, Argentina}       
\centerline{$^{2}$LAFEX, Centro Brasileiro de Pesquisas F{\'\i}sicas,         
                  Rio de Janeiro, Brazil}                                     
\centerline{$^{3}$Universidade do Estado do Rio de Janeiro,                   
                  Rio de Janeiro, Brazil}                                     
\centerline{$^{4}$Institute of High Energy Physics, Beijing,                  
                  People's Republic of China}                                 
\centerline{$^{5}$Universidad de los Andes, Bogot\'{a}, Colombia}             
\centerline{$^{6}$Charles University, Center for Particle Physics,            
                  Prague, Czech Republic}                                     
\centerline{$^{7}$Institute of Physics, Academy of Sciences, Center           
                  for Particle Physics, Prague, Czech Republic}               
\centerline{$^{8}$Universidad San Francisco de Quito, Quito, Ecuador}         
\centerline{$^{9}$Institut des Sciences Nucl\'eaires, IN2P3-CNRS,             
                  Universite de Grenoble 1, Grenoble, France}                 
\centerline{$^{10}$CPPM, IN2P3-CNRS, Universit\'e de la M\'editerran\'ee,     
                  Marseille, France}                                          
\centerline{$^{11}$Laboratoire de l'Acc\'el\'erateur Lin\'eaire,              
                  IN2P3-CNRS, Orsay, France}                                  
\centerline{$^{12}$LPNHE, Universit\'es Paris VI and VII, IN2P3-CNRS,         
                  Paris, France}                                              
\centerline{$^{13}$DAPNIA/Service de Physique des Particules, CEA, Saclay,    
                  France}                                                     
\centerline{$^{14}$Universit{\"a}t Mainz, Institut f{\"u}r Physik,            
                  Mainz, Germany}                                             
\centerline{$^{15}$Panjab University, Chandigarh, India}                      
\centerline{$^{16}$Delhi University, Delhi, India}                            
\centerline{$^{17}$Tata Institute of Fundamental Research, Mumbai, India}     
\centerline{$^{18}$Seoul National University, Seoul, Korea}                   
\centerline{$^{19}$CINVESTAV, Mexico City, Mexico}                            
\centerline{$^{20}$FOM-Institute NIKHEF and University of                     
                  Amsterdam/NIKHEF, Amsterdam, The Netherlands}               
\centerline{$^{21}$University of Nijmegen/NIKHEF, Nijmegen, The               
                  Netherlands}                                                
\centerline{$^{22}$Institute of Nuclear Physics, Krak\'ow, Poland}            
\centerline{$^{23}$Joint Institute for Nuclear Research, Dubna, Russia}       
\centerline{$^{24}$Institute for Theoretical and Experimental Physics,        
                   Moscow, Russia}                                            
\centerline{$^{25}$Moscow State University, Moscow, Russia}                   
\centerline{$^{26}$Institute for High Energy Physics, Protvino, Russia}       
\centerline{$^{27}$Lancaster University, Lancaster, United Kingdom}           
\centerline{$^{28}$Imperial College, London, United Kingdom}                  
\centerline{$^{29}$University of Arizona, Tucson, Arizona 85721}              
\centerline{$^{30}$Lawrence Berkeley National Laboratory and University of    
                  California, Berkeley, California 94720}                     
\centerline{$^{31}$University of California, Davis, California 95616}         
\centerline{$^{32}$California State University, Fresno, California 93740}     
\centerline{$^{33}$University of California, Irvine, California 92697}        
\centerline{$^{34}$University of California, Riverside, California 92521}     
\centerline{$^{35}$Florida State University, Tallahassee, Florida 32306}      
\centerline{$^{36}$Fermi National Accelerator Laboratory, Batavia,            
                   Illinois 60510}                                            
\centerline{$^{37}$University of Illinois at Chicago, Chicago,                
                   Illinois 60607}                                            
\centerline{$^{38}$Northern Illinois University, DeKalb, Illinois 60115}      
\centerline{$^{39}$Northwestern University, Evanston, Illinois 60208}         
\centerline{$^{40}$Indiana University, Bloomington, Indiana 47405}            
\centerline{$^{41}$University of Notre Dame, Notre Dame, Indiana 46556}       
\centerline{$^{42}$Iowa State University, Ames, Iowa 50011}                   
\centerline{$^{43}$University of Kansas, Lawrence, Kansas 66045}              
\centerline{$^{44}$Kansas State University, Manhattan, Kansas 66506}          
\centerline{$^{45}$Louisiana Tech University, Ruston, Louisiana 71272}        
\centerline{$^{46}$University of Maryland, College Park, Maryland 20742}      
\centerline{$^{47}$Boston University, Boston, Massachusetts 02215}            
\centerline{$^{48}$Northeastern University, Boston, Massachusetts 02115}      
\centerline{$^{49}$University of Michigan, Ann Arbor, Michigan 48109}         
\centerline{$^{50}$Michigan State University, East Lansing, Michigan 48824}   
\centerline{$^{51}$University of Nebraska, Lincoln, Nebraska 68588}           
\centerline{$^{52}$Columbia University, New York, New York 10027}             
\centerline{$^{53}$University of Rochester, Rochester, New York 14627}        
\centerline{$^{54}$State University of New York, Stony Brook,                 
                   New York 11794}                                            
\centerline{$^{55}$Brookhaven National Laboratory, Upton, New York 11973}     
\centerline{$^{56}$Langston University, Langston, Oklahoma 73050}             
\centerline{$^{57}$University of Oklahoma, Norman, Oklahoma 73019}            
\centerline{$^{58}$Brown University, Providence, Rhode Island 02912}          
\centerline{$^{59}$University of Texas, Arlington, Texas 76019}               
\centerline{$^{60}$Texas A\&M University, College Station, Texas 77843}       
\centerline{$^{61}$Rice University, Houston, Texas 77005}                     
\centerline{$^{62}$University of Virginia, Charlottesville, Virginia 22901}   
\centerline{$^{63}$University of Washington, Seattle, Washington 98195}       
}                                                                             

\maketitle
%
%
\begin{abstract}
We present the results of a search for leptoquark ($LQ$) pairs in (85.2  
$\pm$ 3.7) pb$^{-1}$ of $p{\bar p}$ collider data collected by the D\O\ 
experiment at the Fermilab Tevatron.  We observe no evidence for leptoquark 
production and set a limit on $\sigma (p{\bar p} \rightarrow LQ\overline{LQ} 
\rightarrow \nu\nu$ + jets) as a function of the mass of the leptoquark 
($m_{LQ}$).  Assuming the decay LQ $\rightarrow \nu q$, we exclude scalar 
leptoquarks for $m_{LQ} <$ 98 GeV/$c^2$, and vector leptoquarks for $m_{LQ} 
<$ 200 GeV/$c^2$ and coupling which produces the minimum cross section, at a 
95\% confidence level. 

\end{abstract}

\newpage
\vfill\eject
\twocolumn
The observed symmetry between the lepton ($l$) and quark ($q$) sectors 
suggests the existence of a force connecting the two that is mediated by
leptoquark ($LQ$) particles that couple directly to both leptons and 
quarks.  Such particles arise naturally as vector \cite{lq} or scalar bosons 
\cite{lqhiggs} in Grand Unified Theories \cite{lq}, as composite particles 
\cite{lqcomp}, as techniparticles \cite{lqtechni}, or as R-parity violating 
supersymmetric particles \cite{rpvio}.
\par 
Leptoquarks would carry both color and fractional electric charge.  They could 
be pair-produced at the Fermilab Tevatron through a virtual gluon ($g$) in the 
strong process $p{\bar p} \rightarrow g \rightarrow LQ\overline{LQ} + X$, with 
a production cross section that, for scalar leptoquarks, is independent of the 
$LQ-q-l$ coupling. For vector leptoquarks, we consider the specific cases of 
the coupling resulting in the minimal cross section ($\sigma _{\rm min}$), 
Minimal Vector coupling (MV), and Yang-Mills coupling (YM) \cite{mc}.   
\par 
Limits from flavor-changing neutral currents imply that leptoquarks of low 
mass ${\cal O}$(TeV) couple only within a single generation \cite{generation}, 
and the decays of leptoquark pairs would therefore be expected to yield one of 
three possible final states:  $l^{\pm}l^{\mp} q\overline{q}$, 
$l^{\pm}\! \! \! \stackrel{(-)}{\nu} \! \! \! q\overline{q}$, and 
$\nu\overline{\nu} q\overline{q}$.  This analysis \cite{thesis} is based on 
the $\nu\overline{\nu} q\overline{q}$ final state, and is sensitive to 
leptoquarks of all three generations.  In a previous study of this final state 
\cite{d01a} with the assumed decay $LQ \rightarrow \nu q$, D\O\ set limits of 
$m_{LQ} >$ 79 GeV/$c^2$ for scalar leptoquarks, and $m_{LQ} >$ 144 GeV/$c^2$, 
159 GeV/$c^2$, and 206 GeV/$c^2$, for vector leptoquarks with couplings that 
correspond to $\sigma _{\rm min}$, MV, and YM couplings, respectively 
\cite{d01a,d01avector}.  The present analysis is based on a factor of 
ten increase in data over the previous analysis.  The CDF collaboration has 
conducted a search for second and third generation leptoquarks, also assuming 
the decay $LQ \rightarrow \nu q$, and set mass limits of 123 (148) GeV/$c^2$ 
for second (third) generation scalar leptoquarks, and 171 (199) GeV/$c^2$ and 
222 (250) GeV/$c^2$ for second (third) generation vector leptoquarks with MV 
and YM couplings, respectively \cite{cdfg23}.  The OPAL collaboration has 
searched $\sqrt{s}$=183 GeV $e^+e^-$ collisions for vector and scalar 
leptoquarks with specific weak isospins and decay modes \cite{OPAL}.  For 
first and second generation scalar leptoquarks with weak isospin of 1 and the 
decay $LQ \rightarrow \nu q$, OPAL has set a mass limit of 84.8 
GeV/$c^2$.  For other values of weak isospin, the mass limit ranges from 71.6 
GeV/$c^2$ to 80.8 GeV/$c^2$.  Our new results extend the range of sensitivity 
of the vector leptoquark searches and the first generation scalar leptoquark 
searches.
\par 
The D\O\ detector \cite{detector} consists of three major subsystems:  an  
inner detector for tracking charged particles; an uranium/liquid-argon 
calorimeter for measuring electromagnetic and hadronic showers; and a muon  
spectrometer.  The inner detector consists of two outer drift chambers 
separately covering the regions $|\eta|<1$ and 1.2 $< |\eta| <$ 2.8, and an
inner drift chamber covering the region $|\eta|<$2.  The calorimeter consists
of three cryostats supplemented with scintillators between the cryostats.  The 
Main Ring beam pipe used to accelerate and inject protons and antiprotons into 
the Tevatron traverses the hadronic region of the calorimeter at 
$\phi$=100$^{\circ}$.  The jets measured with the calorimeter have a 
resolution of approximately $\delta E$=0.8$\sqrt{E}$ ($E$ in GeV).  We measure 
the missing transverse energy (\met) by summing the calorimeter energy 
vectorially in the plane transverse to the beam.  The projection of \met\ on a 
given axis has a resolution of $\delta$\mexy\ =1.08 GeV + 
0.019($\Sigma |E_{x,y}|$) ($E_{x,y}$ in GeV).  
\par 
The event sample for our search is collected with a trigger requiring a jet 
with $E_T >$ 25 GeV, a second jet with $E_T >$ 10 GeV, \met\ $>$ 25 GeV, and 
the azimuthal angle between any jet and \met ($\Delta\phi$(jet, \met)) greater
than 14.3$^{\circ}$.  The trigger does not collect data in the 0.4 seconds 
following Main Ring injection, or within 800 nanoseconds of protons or 
antiprotons passing through the detector.  We remove data affected by 
accelerator noise or detector malfunctions.  The former are identified by 
significant energy measurement in the region surrounding the Main Ring.  The 
latter are identified by recurring energy measurement in a particular region 
of the calorimeter, by energy measurement isolated to a single calorimeter 
cell, and by documented subsystem malfunctions.  The integrated luminosity for 
this sample corresponds to 85.2 $\pm$ 3.7 pb$^{-1}$.  
\par
We select events with well-understood trigger efficiency by requiring at least 
two jets with $E_T >$ 50 GeV, \met\ $>$ 40 GeV, $\Delta\phi$(jet, \met) $ >$ 
30$^{\circ}$, and $\Delta {\cal R} $(jet, jet) $>$ 1.5, where 
${\Delta \cal R}=\sqrt{(\Delta\eta)^2 + (\Delta\phi)^2}$, $\eta$ is the jet 
pseudorapidity, and $\phi$ is the jet azimuthal angle.  Jets are defined as 
the calorimeter energy within a ${\Delta \cal R}=0.5$ cone.  We reduce 
cosmic-ray backgrounds by rejecting events containing jets with little energy 
in the electromagnetic sections of the calorimetry.  Backgrounds arising from 
$W$ or $Z$ boson production are reduced by rejecting events with isolated 
muons or jets with a large fraction of their energy measured in the 
electromagnetic calorimeter.
\par 
The remaining backgrounds in the sample consist of events with jets produced 
in association with a $W$ or a $Z$ boson, and events from top quark and 
multijet production.  We use Monte Carlo generators to simulate the kinematics 
and topologies of events with $W$ or $Z$ bosons or top quarks, and a 
{\sc geant}-based simulation \cite{geant} of the detector to predict the 
acceptance for these events.
\par
The $W$ and $Z$ backgrounds correspond to processes involving only neutrinos 
and jets ($Z$ + 2 jets $\rightarrow \nu\nu$ + 2 jets and $W$ + jet 
$\rightarrow \tau\nu +$ jet, with $\tau \rightarrow $ hadrons + $\nu$), 
processes with undetected charged leptons ($W$ + 2 jets $\rightarrow l^{\pm} 
\nu$ + 2 jets, $Z$ + 2 jets $\rightarrow \mu\mu$ + 2 jets, and $Z$ + jet 
$\rightarrow \tau\tau$ + jet, with one $\tau \rightarrow$ hadrons + $\nu$), 
and processes in which an electron is misidentified as a jet ($W$ + jet 
$\rightarrow e\nu$ + jet and $W$ + jet $\rightarrow \tau\nu$ + jet, with $\tau
\rightarrow e\nu\nu$).  We use the {\sc pythia} Monte Carlo generator 
\cite{pythia} to generate the $W/Z$ + jet processes, and the {\sc vecbos} 
Monte Carlo generator \cite{vecbos} to generate the $W/Z$ + 2 jets 
processes.  We scale the generator cross sections to match the corresponding 
$W/Z$ + jet(s) cross sections measured using decays into electrons.  These 
cross sections were remeasured specifically for this analysis.
\par 
To obtain the background from $t{\bar t}$, $t{\bar b}$, and ${\bar t}b$ 
production, where the top quark decays to an unobserved charged lepton, a 
neutrino, and a jet, we use our measured cross section for $t{\bar t}$ 
production \cite{ttbar}, and the calculated next-to-leading-order cross 
section for the single-top production processes \cite{singlet}.  We use 
the {\sc herwig} Monte Carlo \cite{herwig} program to generate $t{\bar t}$ 
events and the \footnotesize CompHEP \normalsize Monte Carlo \cite{comphep} 
program to generate $t{\bar b}$ and ${\bar t}b$ events. 
\par 
The multijet background arises primarily from a mismeasurement of the 
interaction vertex or of jet energy.  To reduce the number of events with 
mismeasured vertices, we use the central drift chamber (CDC) to associate 
tracks with the two highest $E_T$ jets, if those jets are in the fiducial 
volume of the CDC ($|\eta|\leq$1).  These tracks are used to determine the 
point-of-origin of each jet, which is required to be no further than 15 cm 
from the reconstructed event vertex (the latter is determined from all tracks 
in the event).  The 15 cm value is chosen to maximize the inverse of the 
fractional uncertainty on signal (see below).  We reduce the number of events 
with poorly measured jet energies by requiring that the azimuth $\Delta\phi$ 
between the \met\ vector and the direction of the jet with the second highest 
$E_T$ exceed 60$^{\circ}$.  Table \ref{tbl:initial_cuts} shows the number of 
events remaining in the data after each additional selection criterion.

\begin{table}[ht]
\begin{center}
\begin{tabular}{cc}
Selection criterion                               & \# of Events \\
\hline
  2 Jets + \met\ Trigger                  & 503,557       \\
  No accelerator noise or detector malfunctions &   399,557       \\
  Leading jet $E_T \geq$ 50 GeV           & 236,339       \\
  Second jet $E_T \geq$ 50 GeV            & 86,826       \\
  \met\ $\geq$ 40 GeV                     &  8,996       \\
  $\Delta\phi$(jet,\met) $\geq$ 30$^o$    &  1,567       \\
  $\Delta {\cal R} $(jet, jet) $>$ 1.5    &  1,495  \\
  Jet EM Fraction cuts                    &  1,358  \\
  No isolated muons                       &  1,332  \\
  Leading or second jet $|\eta|\leq$1.0; all jets $|\eta|\leq$4.0&   1,071     \\
  $|$Jet vertex - Primary vertex$|<$ 15 cm  & 401 \\
  $\Delta\phi$(jet 2,\met) $\geq$ 60$^o$  &  231  \\
\end{tabular}
\end{center}
\vskip -0.1in
\caption{The set of criteria imposed on the 2 jets + \met\ data sample and 
the number of events that pass each additional selection criterion. }
\label{tbl:initial_cuts}
\end{table}

To estimate the remaining multijet background in our search sample, we count 
events in which jet-based vertex positions deviate by 15 cm to 50 cm from the 
position of the event vertex.  In events with two central ($|\eta|\leq$1) 
jets, we require both vertices to fall within this range.  We normalize these 
events to the search sample using a multijet-dominated sample 
($\Delta\phi$(jet 2, \met)$< 60^{\circ}$).  The expected multijet background
is:
\large
\begin{center}
$N_{mj}=N_{\Delta\phi > 60^{\circ}}^{15 < z < 50}(\frac{N^{z  < 15}}{N^{15 < z  < 50}})_{\Delta\phi < 60^{\circ}}$
\end{center}  
\normalsize
\noindent 
We choose the upper bound of 50 cm to provide the best match between expected 
background and data for events with \met\ between 30 GeV and 40 GeV, a region
dominated by multijet events.  We predict 162.8 $\pm$ 23.7 multijet and 51.9 
$\pm$ 7.0 $W$, $Z$, and top events in this sample.  We observe 224 events in 
the data.  Changing the vertex threshold to 100 cm increases the multijet 
background prediction by 22\% in this region, which we take as an estimate of 
the systematic error of the method.  Table \ref{tbl:datbd} shows the total 
expected background and the observed number of events for the final 2 jets + 
\met\ data sample. 
\par 
To model the characteristics of leptoquark production, we use scalar  
leptoquark events generated with the {\sc pythia} Monte Carlo program and 
vector leptoquark events generated with the 
\footnotesize CompHEP \normalsize Monte Carlo program.  The cross sections for 
scalar leptoquark production have been calculated to next-to-leading order 
\cite{kraemer}, while those for vector leptoquark production have been 
calculated to leading order \cite{vlqxsec}.  The calculations use a QCD
renormalization and factorization scale of $\mu$=$m_{LQ}$, with theoretical 
uncertainties estimated by changing the scale to $\mu$=$m_{LQ}$/2 and 
$\mu$=2$m_{LQ}$.  For scalar leptoquarks we use the smaller predicted cross 
section ($\mu$=2$m_{LQ}$) for determining the mass limits on $LQ$'s.
\par
\begin{figure}[!htbp]  
\begin{minipage}[htb]{4.3cm} 
\epsfysize = 4.3cm  
\centerline{\epsffile{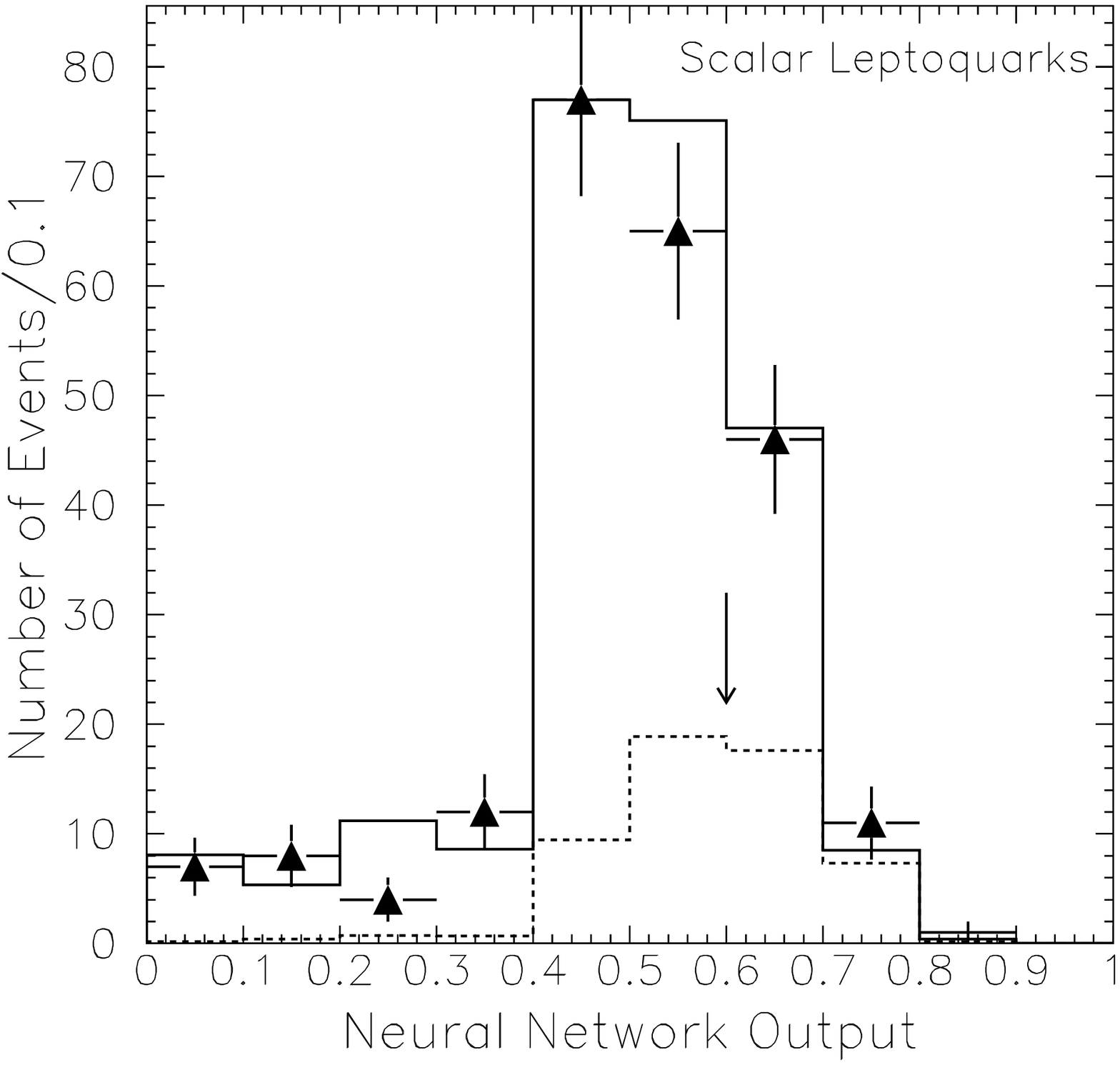}} 
\end{minipage} 
\hskip -0.1cm
\begin{minipage}[htb]{4.3cm} 
\epsfysize = 4.3cm  
\centerline{\epsffile{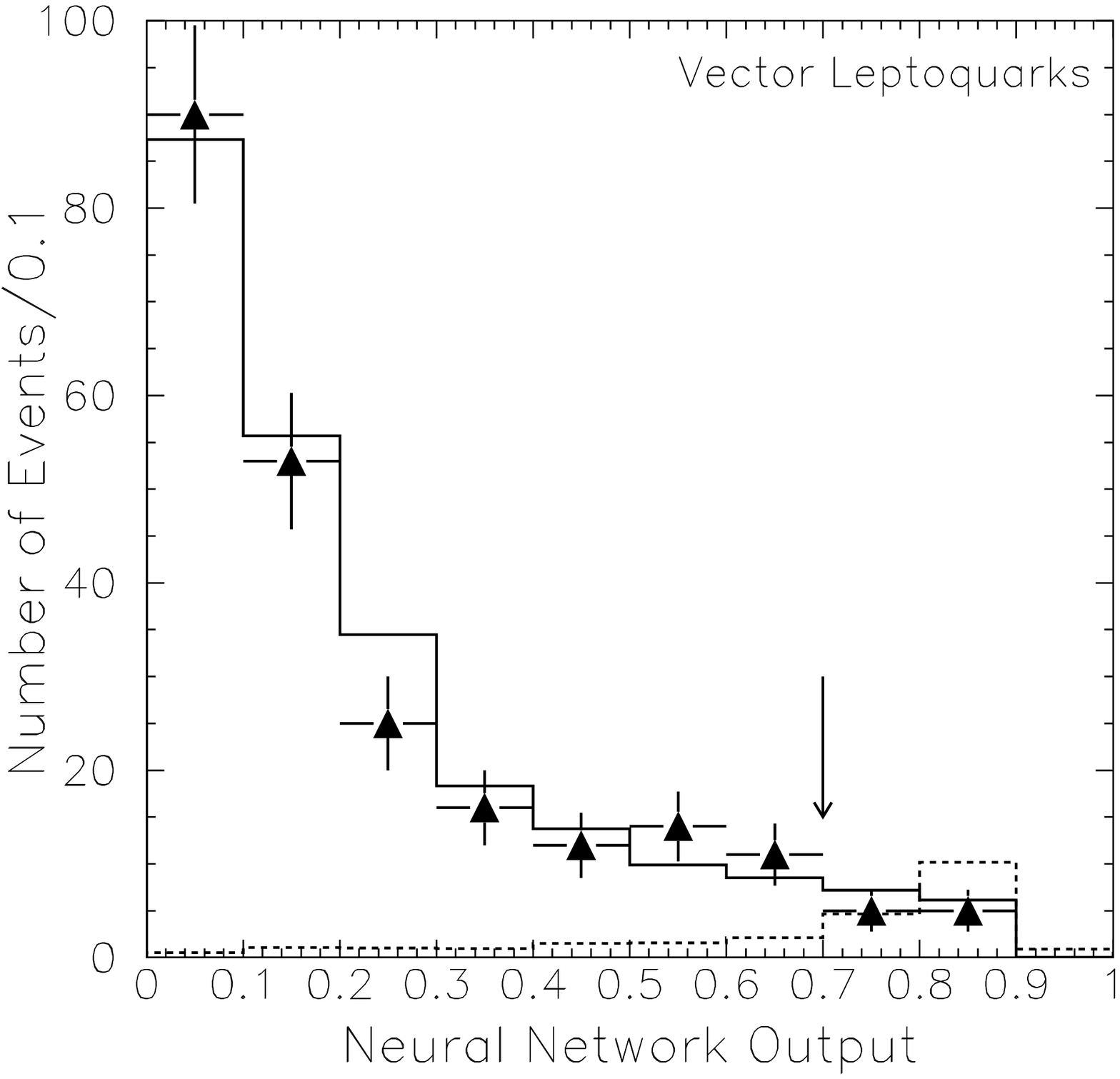}} 
\end{minipage} 
\caption{The neural network output for data (points), for background (solid
histogram), and for leptoquarks (dashed histogram).  The optimization is for 
100 GeV/$c^2$ scalar leptoquarks (left) and 200 GeV/$c^2$ vector leptoquarks 
with Minimal Vector coupling (right).  We remove events to the left of the 
arrows.} 
\label{fig:nnout} 
\end{figure}

 
Failure to observe any hypothetical signal at 95\% confidence level (C.L.)
corresponds approximately to a downward fluctuation of that signal by two 
standard deviations.  Hence, to increase the sensitivity of our search for the
production of leptoquarks that decay to $\nu q$, we search for leptoquarks 
that would produce excesses of approximately two standard deviations.  We 
separately optimize our selection criteria for the production of 100 GeV/$c^2$ 
scalar leptoquarks and for 200 GeV/$c^2$ vector leptoquarks with Minimal 
Vector coupling.  Other choices of leptoquark masses do not significantly 
affect our results.  We use the {\sc jetnet} \cite{jetnet} neural network 
program to isolate regions of significant leptoquark production, with \met\ 
and $\Delta\phi$(jet, jet) as inputs for scalar leptoquarks, and \met\ and the 
$E_T$ of the jet with the second highest $E_T$ as inputs for vector 
leptoquarks.  The values of the neural network output variables and the 
thresholds for these masses are shown in Fig. \ref{fig:nnout}.  The thresholds 
are chosen to maximize the quantity:
\large
\begin{center}
$\frac{N_{lq}}{\sqrt{N_{lq}+N_{\rm back}+(\Delta N_{lq})^2+(\Delta N_{\rm back})^2}}$,
\end{center}  
\normalsize
\noindent 
where $N_{lq}$ and $N_{\rm back}$ are the number of signal and background
events, respectively, and $\Delta N_{lq}$ and $\Delta N_{\rm back}$ are their
associated uncertainties.  This quantity reflects the inverse of the 
fractional uncertainty on signal.  After applying these thresholds, we expect
56.0$^{+8.1} \! \! \! \! \! \! \! \! \! \! \! _{-8.2}$ events and observe
58 events for the scalar leptoquark optimization, and expect
13.3$^{+2.8} \! \! \! \! \! \! \! \! \! \! \!  _{-2.6}$ events and observe
10 events for the vector leptoquark optimization.


\begin{table}[hptb] 
\begin{center} 
\begin{tabular}{cc} 
Type of Events & \# of Events \\ 
\hline
 Multijet & 58.8 $\pm$ 14.1 $\pm$ 12.9\\ 
($W \rightarrow e\nu$) + jet & 51.9 $\pm$ 7.0 $^{+13.7} \! \! \! \! \! \! \! \! \! \! \! \! \! _{-8.9}$ \\ 
($W \rightarrow \tau\nu$) + jet & 46.3 $\pm$ 5.0 $^{+8.9} \! \! \! \! \! \! \! \! \! \! _{-7.7}$ \\ 
($Z \rightarrow \nu\nu$) + 2 jets & 36.1 $\pm$ 7.7 $^{+9.0} \! \! \! \! \! \! \! \! \! \! \! _{-5.5}$ \\ 
($W \rightarrow \mu\nu$) + 2 jets & 18.7 $\pm$ 3.5 $^{+4.2} \! \! \! \! \! \! \! \! \! \! _{-3.7}$ \\ 
$t{\bar t} \rightarrow$ $l^{\pm}\nu$ + 4 jets & 10.6 $\pm$ 2.0 $\pm$ 2.3 \\ 
($W \rightarrow e\nu$) + 2 jets & 8.3 $\pm$ 2.5 $^{+2.0} \! \! \! \! \! \! \! \! \! \! _{-2.5}$ \\ 
($W \rightarrow \tau\nu$) + 2 jets & 5.6 $\pm$ 1.7 $^{+1.4} \! \! \! \! \! \! \! \! \! \! _{-0.8}$ \\ 
$tb \rightarrow$ $l^{\pm}\nu$ + 2 jets & 2.0 $\pm$ 0.3 $\pm$ 0.2 \\  
($Z \rightarrow \tau\tau$) + jet & 2.0 $\pm$ 0.4 $^{+0.6} \! \! \! \! \! \! \! \! \! \! _{-0.3}$ \\ 
($Z \rightarrow \mu\mu$) + 2 jets & 1.7 $\pm$ 0.4 $^{+0.4} \! \! \! \! \! \! \! \! \! \! _{-0.3}$ \\ 
$~~~~~~~~~~~~~$ Total background $~~~~~~~~~~~~~$& 242.0 $\pm$ 18.9 $^{+23.3} \! \! \! \! \! \! \! \! \! \! \! \! \! _{-19.0}$ \\ 
 Data & 231 \\ 
\end{tabular} 
\end{center} 
\caption{The expected and observed numbers of events in the final 2 jets + 
\met\ sample.} 
\label{tbl:datbd} 
\end{table} 

\begin{figure}[htb] 
\begin{center}
\begin{minipage}[htb]{5.5cm} 
\epsfysize = 5.5cm  
\centerline{\epsffile{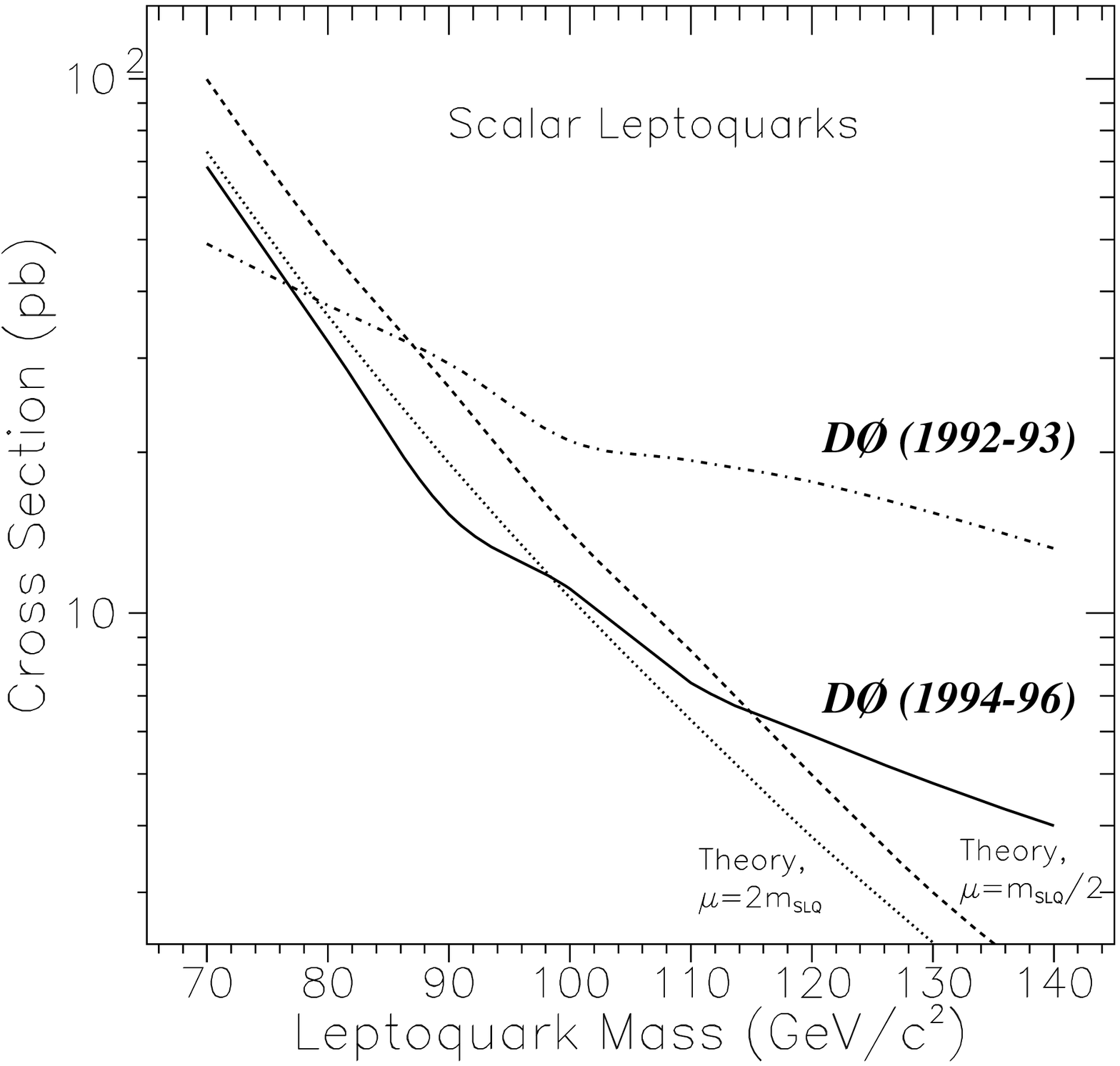}} 
\end{minipage} 
\begin{minipage}[htb]{5.5cm} 
\epsfysize = 5.5cm  
\centerline{\epsffile{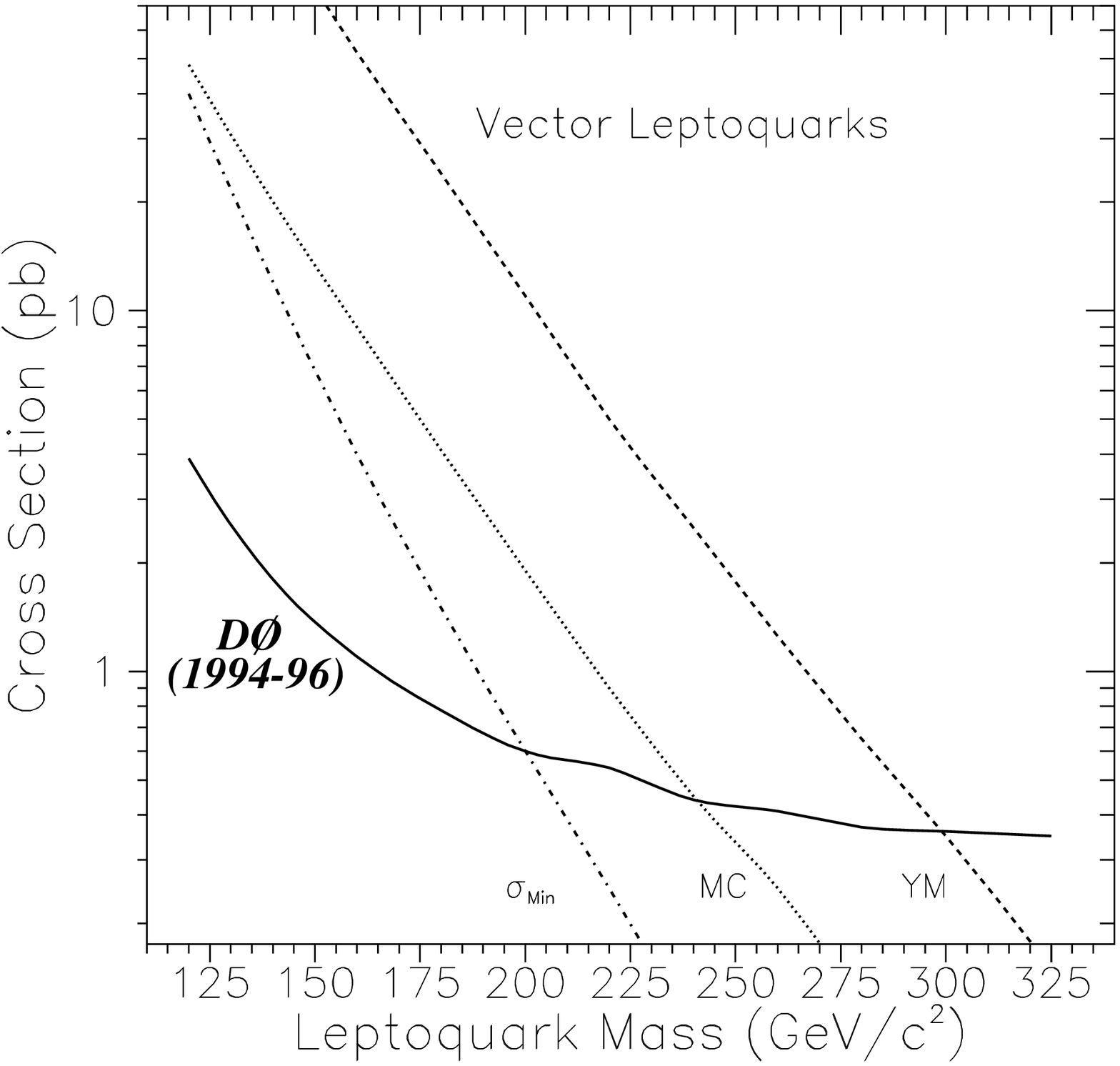}} 
\end{minipage} 
\caption{Limits on cross section at 95\% confidence level, as a function of 
leptoquark mass, for scalar (top) and vector (bottom) leptoquarks, and 
different theoretical predictions. We assume the $LQ$ decays exclusively to 
$\nu q$.  The theoretical predictions correspond to the production of 
leptoquarks of a single generation, while the experimental limit corresponds 
to the sum of contributions from leptoquarks of all three generations. }
\label{fig:lqlim} 
\end{center}
\end{figure}

\begin{figure}[htb] 
\begin{center}
\begin{minipage}[htb]{7.0cm} 
\epsfysize = 7.0cm  
\centerline{\epsffile{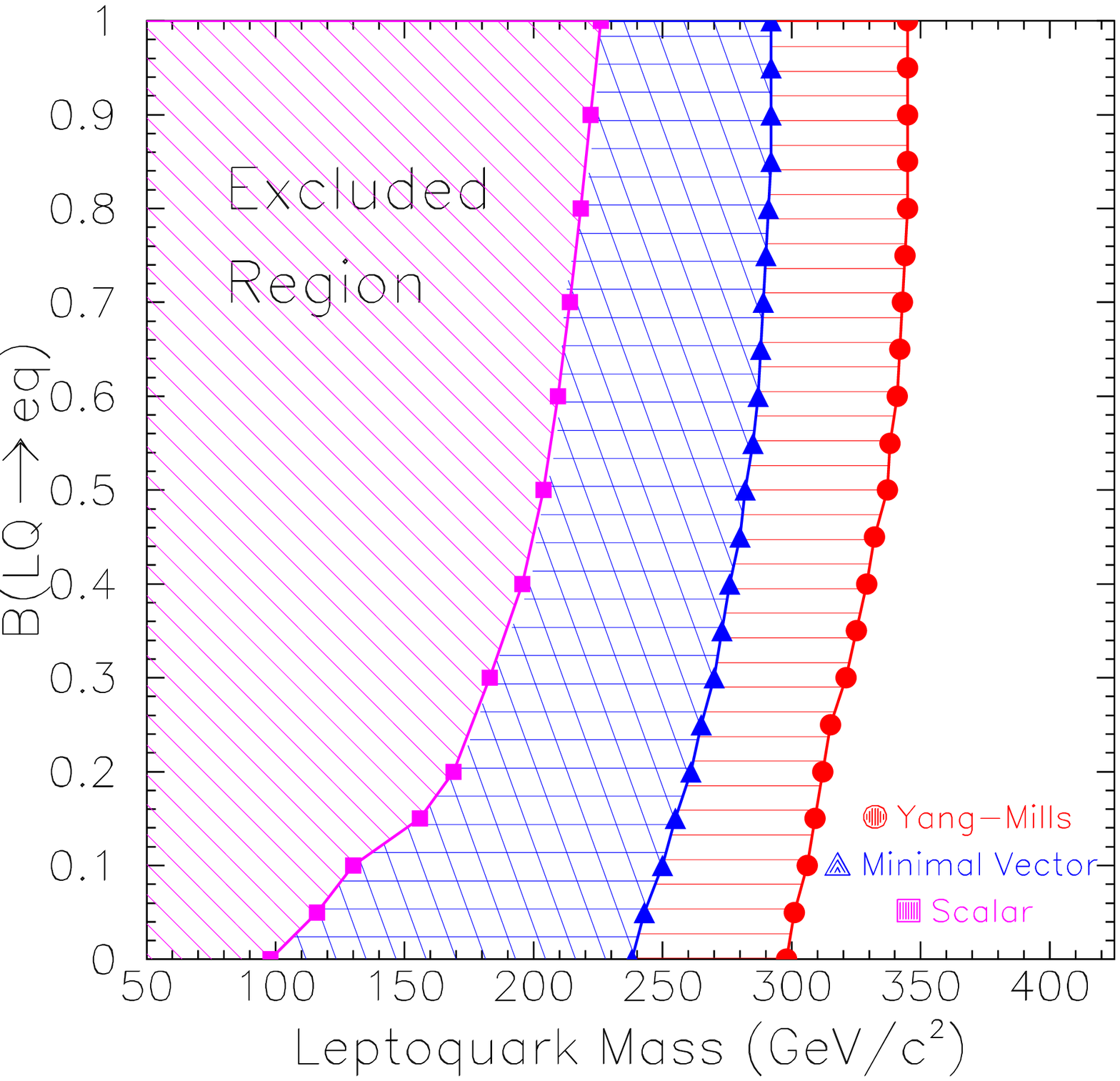}} 
\end{minipage} 
\begin{minipage}[htb]{7.0cm} 
\epsfysize = 7.0cm  
\centerline{\epsffile{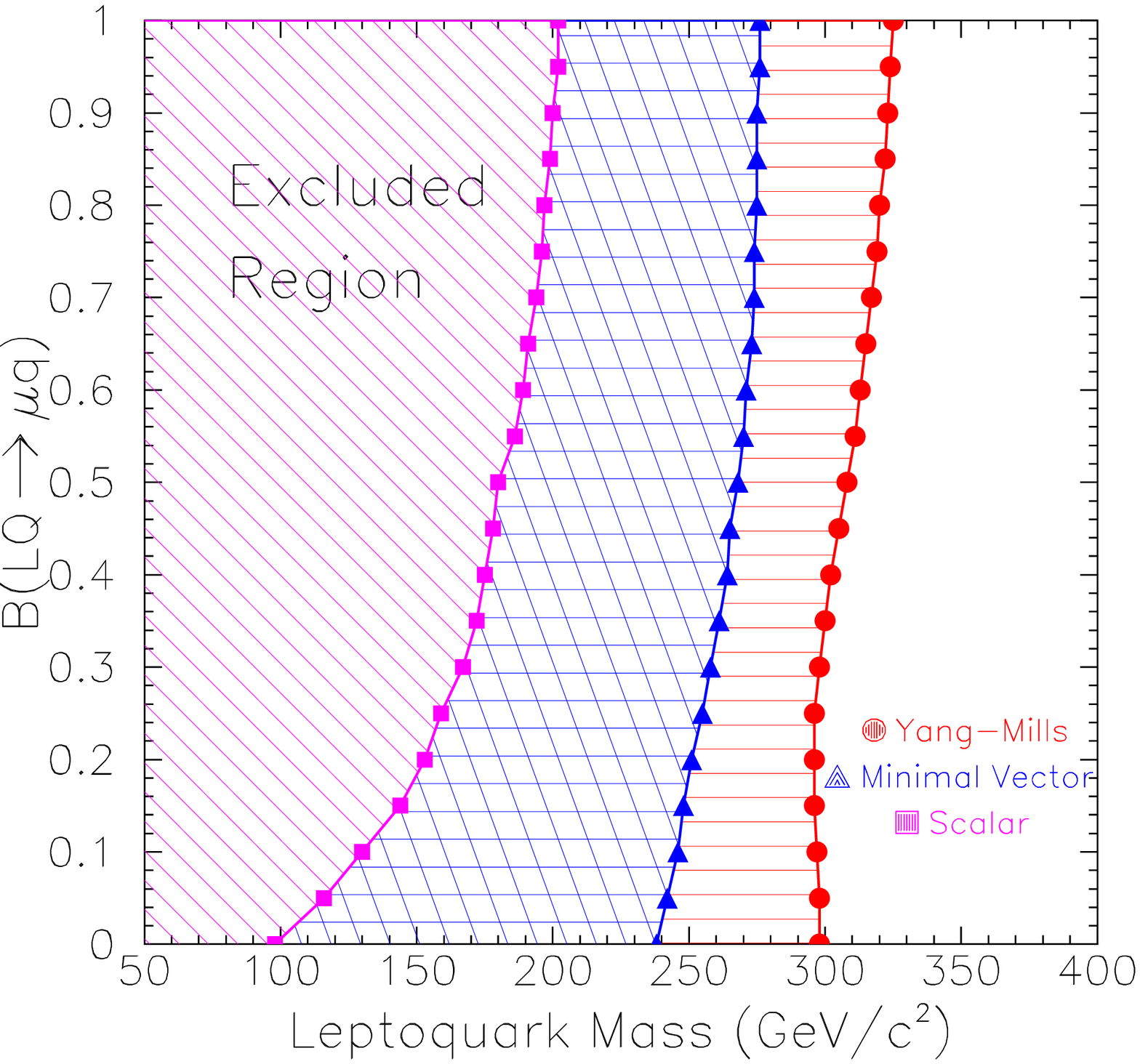}} 
\end{minipage} 
\caption{The region of ${\cal B}$($LQ \rightarrow l^{\pm}q$) vs. mass for 
first-generation (top) and second-generation (bottom) leptoquarks excluded by 
D\O.} 
\label{fig:brvmass} 
\end{center}
\end{figure}

After applying the optimal thresholds, we find that the observed number of 
events is consistent with the expected background, and that, consequently, we 
have no evidence for leptoquark production.  This null result yields the 
95\% C.L. upper limit on cross section (Fig. \ref{fig:lqlim}) as a 
function of leptoquark mass.  We calculate the limit using a Bayesian method 
\cite{bayes} with a flat prior for the signal and Gaussian priors for 
background and acceptance uncertainties.  The equivalent limits on mass are 98 
GeV/$c^2$ for scalar leptoquarks, and 200 GeV/$c^2$, 238 GeV/$c^2$, and 
298 GeV/$c^2$ for vector leptoquarks with couplings corresponding to the 
minimum cross section $\sigma _{\rm min}$, Minimal Vector coupling, and 
Yang-Mills coupling, respectively.  We summarize the D\O\ mass limits as a 
function of the branching fraction ${\cal B}$($LQ \rightarrow l^{\pm}q$) for 
first \cite{d01a} and second \cite{d0g2} generation leptoquarks in Fig. 
\ref{fig:brvmass}.  These limits combine searches using the final states 
$l^{\pm}l^{\mp} q\overline{q}$, $l^{\pm}\! \! \! \stackrel{(-)}{\nu} \! \! \! 
q\overline{q}$, and $\nu\overline{\nu} q\overline{q}$.  We note that the gap 
at small values of ${\cal B}$($LQ \rightarrow l^{\pm}q$) in previous analyses 
has been filled as a result of this investigation.


We thank the staffs at Fermilab and collaborating institutions, 
and acknowledge support from the 
Department of Energy and National Science Foundation (USA),  
Commissariat  \` a L'Energie Atomique and 
CNRS/Institut National de Physique Nucl\'eaire et 
de Physique des Particules (France), 
Ministry for Science and Technology and Ministry for Atomic 
   Energy (Russia),
CAPES and CNPq (Brazil),
Departments of Atomic Energy and Science and Education (India),
Colciencias (Colombia),
CONACyT (Mexico),
Ministry of Education and KOSEF (Korea),
CONICET and UBACyT (Argentina),
The Foundation for Fundamental Research on Matter (The Netherlands),
PPARC (United Kingdom),
Ministry of Education (Czech Republic),
and the A.P.~Sloan Foundation.


\begin{references} 
%
\bibitem[*]{lehner}
Visitor from University of Zurich, Zurich, Switzerland.
\bibitem[\dag]{przybycien}
Visitor from Institute of Nuclear Physics, Krakow, Poland.
%
\vskip 0.25cm
\bibitem{lq}  
H. Georgi and S. Glashow, Phys. Rev.  
Lett. \bf 32\rm , 438 (1974); J.C. Pati and   
A. Salam, Phys. Rev. D \bf 10\rm , 275 (1974).  
%
\bibitem{lqhiggs}  
P.H. Frampton, Mod. Phys. Lett. A \bf 7\rm,   
\rm 559 (1992).    
%
\bibitem{lqcomp}  
J. L. Hewett and T. G. Rizzo, Phys. Rep. \bf 183\rm,   
193 (1989); E. Accomando \em et al.\rm , Phys. Rep.   
\bf 299\rm, 1 (1998).   
%
\bibitem{lqtechni}   
E. Eichten and K. Lane, hep-ph/9609297; hep-ph/9609298, Proceedings of 1996
Snowmass Summer Study.   
%
\bibitem{rpvio}  
D. Choudhury and S. Raychaudhuri, Phys. Lett. B \bf 401\rm ,  
367 (1997); G. Altarelli \em et al.\rm , Nucl. Phys. B \bf 506\rm,   
3 (1997).  
%
\bibitem{mc}   
\rm J. Bl$\rm\ddot{u}$mlein, E. Boos, and A. Kryukov,   
Z. Phys. C \bf 76\rm, 137 (1997).   
%
\bibitem{generation}  
H.-U. Bengtsson, W.-S. Hou, A. Soni, and D. H. Stork,  
Phys. Rev. Lett. \bf 55\rm , 2762 (1985).  
%
\bibitem{thesis} 
C. Hays, Ph.D. thesis, Columbia University, 
2001 (unpublished). 
%
\bibitem{d01a}  
B. Abbott \em et al.\rm (D\O\ Collaboration), Phys. Rev. Lett. \bf 80\rm , 
2051 (1998).
%
\bibitem{d01avector}  
V. M. Abazov \em et al.\rm (D\O\ Collaboration), hep-ex/0105072.
%
\bibitem{cdfg23}  
T. Affolder {\it et al.}, Phys.   
Rev. Lett. \bf 85\rm , 2056 (2000).  
%
\bibitem{OPAL}  
G. Abbiendi {\it et al.}, Eur. Phys. J. C \bf 13\rm , 15 (2000). 
%
\bibitem{detector}  
B. Abbott {\it et al.} (D\O\ Collaboration), Nucl. Instrum. Methods Phys. Res. 
A \bf 338\rm , 185 (1994).  
%
\bibitem{geant}  
R. Brun and F. Carminati, CERN Program Library Long Writeup, W5013, 1993 
(unpublished).  We used version 3.15.
%
\bibitem{pythia}  
T. Sj$\rm\ddot{o}$strand, Comput. Phys. Commun. \bf 82\rm , 74 (1994).  We used
version 6.127.  
%
\bibitem{vecbos}  
F.A. Berends {\it et al.}, Nucl. Phys. B\bf 357\rm , 32 (1991).  
%
\bibitem{ttbar}  
B. Abbott {\it et al.} (D\O\ Collaboration), Phys. Rev. D \bf 60\rm , 012001 
(1999).  
%
\bibitem{singlet}  
M.C. Smith and S. Willenbrock, Phys. Rev. D \bf 54\rm , 6696 (1996); T. 
Stelzer, Z. Sullivan, and  S. Willenbrock, Phys. Rev. D \bf 56\rm , 5919 
(1997); {\sl ibid.} \bf 58\rm , 094021 (1998); V. M. Abazov {\it et al.}, 
Phys. Lett. B \bf 517\rm , 282 (2001).
%
\bibitem{herwig}  
G.~Marchesini {\it et al.}, Comput. Phys. Commun. {\bf 67}, 465   
(1992).  
%
\bibitem{comphep}  
A. Pukhov {\it et al.}, hep-ph/9908288.  We used version 3.0.  
%
\bibitem{kraemer}   
\rm M. Kr$\rm\ddot{a}$mer {\it et al.},   Phys. Rev. Lett. \bf 79\rm,   
\rm 341 (1997).   
%
\bibitem{vlqxsec}   
\rm J. Bl$\rm\ddot{u}$mlein, E. Boos, and A. Kryukov,   
hep-ph/9811271.   
%
\bibitem{jetnet}  
C. Peterson, T. R$\rm\ddot{o}$gnvaldsson, and L.   
L$\rm\ddot{o}$nnblad, Comput. Phys. Commun. {\bf 81}, 185 (1994).  We used 
version 3.4.  
%
\bibitem{bayes}  
I. Bertram {\it et al.}, Fermilab-TM-2104 (unpublished).
%
\bibitem{d0g2}   
B. Abbott {\it et al.} (D\O\ Collaboration), Phys. Rev. Lett. {\bf 84}, 2088 
(2000).
%
\end{references}
\end{document}